\documentclass[twocolumn,aps,prb,floats,showpacs,preprintnumbers,amsmath,amssymb,amsart]{revtex4}
\usepackage{graphicx}
\usepackage{bm}
\usepackage{epsfig}
\input{epsf}
\usepackage{color}
\usepackage{braket}
\newcommand{\chb}{\textcolor{blue}}

\usepackage{hyperref}
\hypersetup{
colorlinks=true,final=true,
        linkcolor=blue,
        citecolor=blue,
        filecolor=blue,
        urlcolor=blue}

\begin{document}

\title{Defect-induced modulation of magnetic, electronic and optical properties of La$_2$CoMnO$_6$ double perovskite oxide}

\author{Anasua Khan$^1$, Swastika Chatterjee$^2$\footnote{swastika@iiserkol.ac.in}, T.K. Nath$^1$ and A. Taraphder$^{1,3}$}
\affiliation{$^1$Department of Physics, Indian Institute of Technology, Kharagpur 721302, India.\\
            $^2$Department of Earth Sciences, Indian Institute of Science Education and Research, Kolkata 741246, India.\\
      $^3$Centre for Theoretical Studies, Indian Institute of Technology, Kharagpur 721302, India.}

\date{\today}                                         

\begin{abstract}

Electron- and hole-doped La$_2$CoMnO$_6$(LCMO) are investigated using first principles DFT calculations. Hole and electron doping are achieved respectively by introducing Sr$^{2+}$ at La$^{3+}$ sites and by inducing O-site vacancies in LCMO. Electronic structure calculations suggest that hole doping alters the charge and valence state of Co ions, whereas electron doping influences the Mn ions. Introduction of defects is found to enhance antisite disorder(ASD) at Co/Mn site, which is expected to influence the magnetodielectric properties of the system. Our calculations suggest that while ASD and/or hole doping induces half-metallicity in LCMO, electron doping restores its insulating state. Mean-field calculations performed using exchange constants obtained by mapping DFT total energies of different collinear spin configurations onto the Ising Hamiltonian find that defects tend to reduce the Curie temperature ($T_C$). Interestingly, the calculated linear optical properties of the system suggest that the material becomes optically active with high values of birefringence in the presence of defects, a property that is highly sought-after in the optical communications and laser industry.

\end{abstract}

\maketitle{}
\section{Introduction} 
Largely because of their fascinating properties that hold a substantial promise for a wide range of device application, research in multifunctional materials has come to the fore in the last two decades. For example, magnetoelectric multiferroic materials SeCoO$_3$, TbMnO$_3$, CdCr$_2$S$_4$, EuO and CuO\cite{1,2,3} have been extensively studied because of their potential applicability in multifunctional devices including sensors\cite{4}, memories\cite{5,6}, magnetic recording readers\cite{7}, transformers and energy harvesting devices\cite{8}.
However, their application in real devices is greatly restricted as their Curie temperatures are well below the room temperature. In this context, transition metal-based double perovskite oxides are of special interest as their Curie temperature lies well above the room temperature. In particular, Co-based perovskites and double perovskites have drawn much attention due to their rich physics arising from the additional spin degree of freedom of Co, resulting in the introduction of several interesting physical phenomena such as large magnetoresistance, insulator-metal transition, Hall effect, orbital ordering and charge ordering\cite{9,10,11,12,13}. For example, the non-magnetic insulating ground state of LaCoO$_3$, with Co in its low spin configuration (LS, S=0, $t_{2g}^6e_{g}^0$), is known to switch to magnetic intermediate spin state (IS, S=1, $t_{2g}^5e_g^1$)\cite{14,15} or high spin state (HS, S=2, $t_{2g}^4e_g^2$)\cite{16,17,18} as temperature increases. LCMO has been studied extensively both in bulk and thin film forms. Initially there was a confusion regarding the nature of the precise valence, spin and magnetic state of Co and Mn. However, later studies from X-ray absorption spectroscopy (XAS)\cite{19} and X-ray absorption near-edge spectroscopy (XANES)\cite{20}confirm that in the ordered state, Co and Mn are in their high spin (+2) and (+4) valence states respectively. Whereas in the disordered phase, both the transition metal ions take up identical charge (+3 valence state) state with Co being in the intermediate spin state and Mn in the high spin state\cite{21,22,23,24}. The perfectly ordered arrangement of Co/Mn in LCMO exhibits a high temperature ferromagnetic transition at T$_c$ = 240 K. A disordered arrangement of Co/Mn, on the other hand, shows an additional low temperature ferromagnetic transition  at T$_c$ =150 K\cite{25}.

The physical and chemical properties of double perovskite oxides become all the more interesting when its A-site is partially doped by a cation with a different ionic radius and/or charge state compared to the host ion.
For example, structural transitions are known to occur as a result of substitution of Sr$^{2+}$ ion by La$^{3+}$ ion in Sr$_2$FeMoO$_6$\cite{A-site3}. Moreover, it has been observed, both theoretically and experimentally, that A-site doping has a significant impact on the valence and the magnetic moment of transition metal ions, which often leads to dramatic changes in magnetic, orbital and transport properties of the material\cite{A-site0,A-site1,A-site2,atpg,BFCO0,BFCO1,BFCO2,BFCO3,BFCO4}.
In order to explore such possibilities in LCMO, we have performed a comprehensive first-principles study of fractional doping of Sr$^{2+}$ at its La$^{3+}$ site. We have particularly addressed the case of LaSrCoMnO$_6$ (LSCMO), where 50\% of the La$^{3+}$ ions are replaced by Sr$^{2+}$\cite{Sr1,Sr2,Sr3,Sr4,Sr5}. It is well known that anti-site disorder, defined as imperfections created by misplacement of the transition metal ions (compared to the perfectly ordered double  perovskite crystal structure), occurs naturally during preparation and plays an important role in determining the physical properties. Experimentally, antisite disorder in LCMO is known to give rise to antiferromagnetic clusters in an otherwise ferromagnetic matrix. Such antiferromagnetic clusters often lead to spin-glass state and metamagnetic transitions\cite{A,B,C,D}. We have, therefore, investigated the probability of occurrence of anti-site disorder in the presence and absence of Sr-doping and also its effect on the physical properties of the system. Additionally, O-site vacancies can easily be incorporated in these oxides by modulating the oxygen fugacity of the environment in which the sample is being prepared\cite{21}. Such point defects, though very few in number, can have profound effect on the properties of the system. For example, the half metallicity of bulk samples have been found to be destroyed in the presence of oxygen vacancies\cite{half metal}. Whereas, in the case of LCMO\cite{21}, a new ferromagnetic phase mediated by vibronic super-exchange interaction\cite{vibronic} appears at lower temperatures in the presence of oxygen vacancy. We have, therefore, also investigated the effect of O-site vacancy on the structural, electronic, magnetic and optical properties and also on the probability of occurrence of anti-site disorder in LCMO and LSCMO. Of particular interest to us is the effect of defects on the optical properties of the material, an avenue that remains mostly unexplored in the case of LCMO. Defects may trigger optical anisotropy, making these materials potential candidate for application in optical isolators, circulators, polarization beam splitters and photo storage devices\cite{optic0,optic1,optic2}.      

\section{Computational details} 
All the theoretical calculations presented here have been performed using density functional theory (DFT) within the framework of generalized gradient approximation(GGA)\cite{28}. For our DFT calculations, we have considered a combination of two different methods, namely a plane-wave based pseudopotential method as implemented in the \textit{Vienna ab-initio simulation package} (VASP)\cite{29,vasp} and full potential linear augmented plane wave (FPLAPW) as implemented in WIEN2k code\cite{30}. The plane-wave-based VASP method was used for structural optimization, while the electronic structures in the energy-minimized crystal structures were calculated using both FPLAPW and pseudopotential method. The accuracy of the electronic structure calculations within the scheme of the two methods have been checked with respect to each other. In the plane wave based pseudopotential calculations, reciprocal space integration was carried out with a $k$ mesh of 8$\times$8$\times$4. We have used projector augmented wave (PAW) potentials and the wave functions were expanded in the plane-wave basis with a kinetic-energy cutoff of 700 eV. The ionic positions have been relaxed using a conjugate-gradient algorithm, until the Helmann-Feynman forces become less than 0.005 eV/\r{A}. The spin-orbit coupling was taken into account self-consistently. To include the strong correlation effects in the 3$d$ electrons of Co and Mn, we used the spin-polarized GGA plus Hubbard U (GGA + U)\cite{31} method, as in the Dudarev$'$s implementation\cite{32}. The nominal value of U$_{eff}$ (U-J = 3 eV) is included at Co and Mn site in both method\cite{hubbard_U,hubbard_U1}. In the full potential linear augmented plane wave, the muffin-tin radii are chosen to be 2.20, 2.00, 1.85, 1.96, and 1.45 a.u. for La, Sr, Co, Mn and O atoms, respectively. We have taken L$_{max}$=12 for spherical harmonic expansion of the wave function inside the sphere and G$_{max}$=14 for the charge Fourier expansion. We have set the basis set parameter R$_{mt}$K$_{max}$=7.00 for the plane wave expansion where the R$_{max}$ is smallest muffin-tin radius and largest K vector is K$_{max}$. The energy convergence criteria for self-consistence calculations are set to 10$^{-5}$ eV. In the calculations spin-orbit coupling was included in scalar relativistic form as a perturbation to the original Hamiltonian. 

The frequency dependent dielectric function calculations in the presence and absence of vacancies, we used the equivalent to 1400 k-points in entire Brillouin Zone by OPTIC code\cite{optic} as implement in the all-electron WIEN2k method\cite{30}.  
\section{Results}
\subsection{Crystal Structure} 
\subsubsection{In the absence of vacancy}

\begin{figure*}[htp]
\centering
\includegraphics[scale=.30]{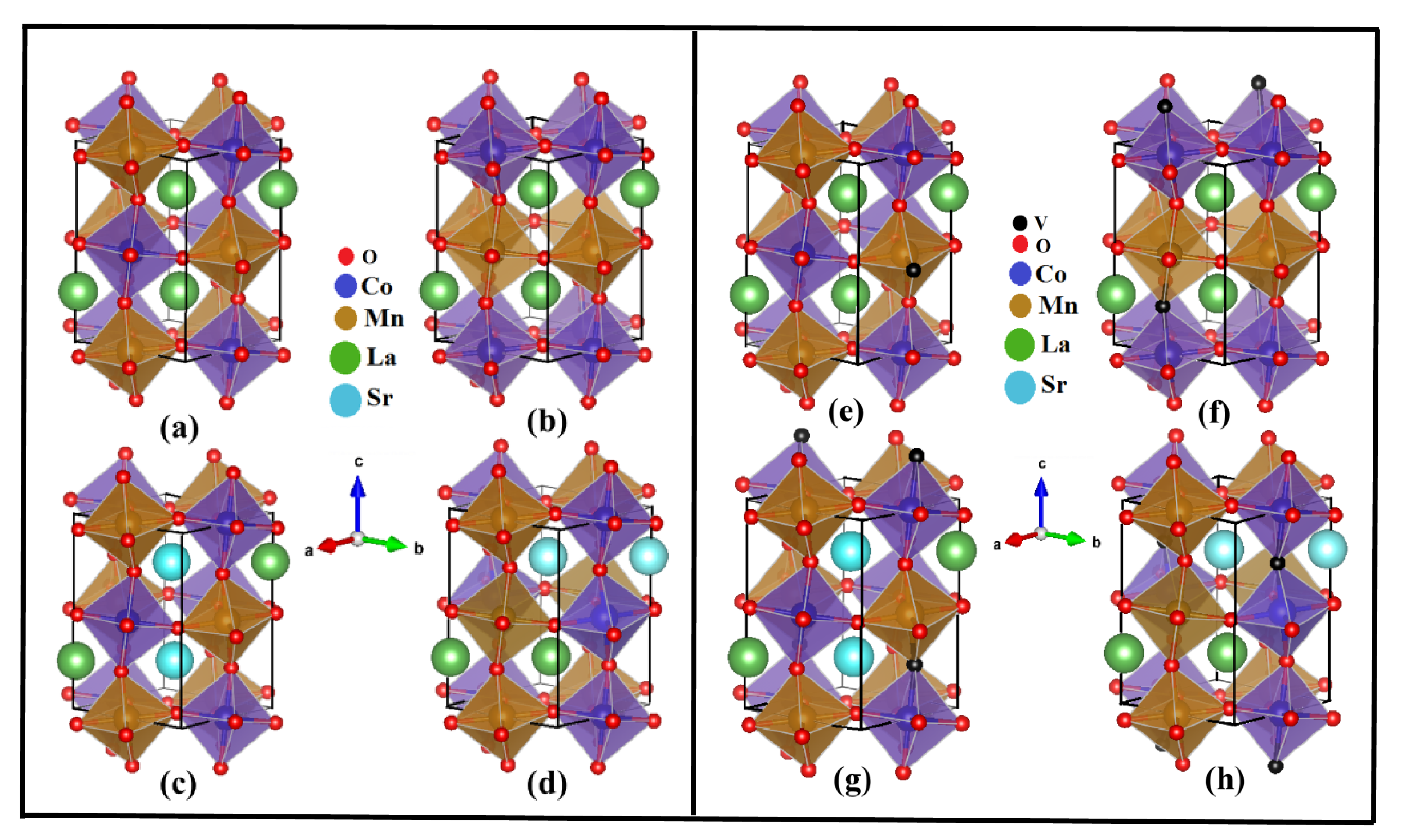} \\ 
\caption{Lowest energy crystal structure of (a) ordered LCMO, (b) disordered LCMO, (c) Sr doped ordered LCMO and (d) Sr doped disordered LCMO, (e) ordered LCMO with O-site vacancy, (f) disordered LCMO with O-site vacancy, (g) Sr doped ordered LCMO with O-site vacancy and (h) Sr doped disordered LCMO with O-site vacancy. Here the centre of the blue and yellow octahedra are occupied by Co and Mn respectively. The La and Sr atoms, represented by green and light blue large balls sit in the hollow formed by MnO$_6$ and CoO$_6$ octahedra. The vacancy is denoted by small black ball.}
\end{figure*}
LCMO is known to crystallize in the monoclinic space group, \textit{P2$_1$/n}. The unit cell contains two formula units, such that there are 4 La, 2 Co, 2 Mn and 12 O atoms per unit cell. Starting from the experimentally obtained crystal structure\cite{coordinate}, we have optimized the lattice parameters and atomic positions of the LCMO to obtain the ground state crystal structure. FIG. 1(a) shows the ordered arrangement of pure LCMO. Here every CoO$_6$ octahedra shares its corners with six MnO$_6$ octahedra. Any other type of arrangement of Co and Mn will result in antisite disorder. Further details may be obtained from the Supplemental Material\cite{supplemental}(SM). We find that the energy difference between the ordered structure and the energetically most favoured disordered (FIG. 1(b)) structure is 0.22 eV/$f.u$. We next consider the effect of Sr doping at La sites of LCMO. In the presence of Sr doping, a new rhombohedral phase ($R\overline{3}c$ ) is known to crystalize\cite{R3c}, apart from the dominant monoclinic \textit{P$2_1/n$} phase. In the monoclinic phase, Sr can be doped in three different ways (refer to S2 of SM\cite{supplemental}). For each of these three atomic arrangements of Sr, the Co and Mn ions can take up either the ordered configuration, or two disordered configurations as already explained. We have calculated the total energy for all these configurations. We find that the ordered structure always possesses the lowest energy, i.e, it is the most stable configuration even in the presence of Sr doping. However, the difference between the lowest energy ordered structure FIG. 1(c) and the lowest energy disordered structure FIG. 1(d) is found to decrease significantly with Sr-doping, the difference being 0.11 eV/$f.u$. From FIGs. 1(c) and 1(d) it may seem that the distribution of La/Sr ions in the lattice depends upon the Co/Mn ordering. However, it may be noted that total energy calculations suggest that this dependence though present, is very weak. The energy difference between two configurations with similar Co/Mn arrangement and different La/Sr arrangement is low, of the order of few meV, which is an order of magnitude lower than the energy difference brought in by disorder at Co/Mn sites. The optimized ground state crystal structure data have been included in TABLE S1 of the SM\cite{supplemental}. We have also seen the effect of Sr-doping on the minor rhombohedral phase, the details can be found in the SM\cite{supplemental}. We conclude from total energy calculations that with Sr doping at La sites, disorder becomes favourable as compared to when there was no Sr doping. The energy difference between the ordered and disordered structure of LCMO and LSCMO in both monoclinic and rhombohedral phases are presented in TABLE I.
\begin{table*}[t]
\caption{Calculated energy difference between ordered and disordered structures of LCMO and LSCMO with and without vacancy.}
\begin{tabular}{| l | c | c |}
\hline
\hline
&\chb{without vacancy}&\chb{with vacancy}\\
  & $\Delta$E = E$_{disordered}$ - E$_{ordered}$ \bf({eV/f.u.})&$\Delta$E = E$_{disordered}$ - E$_{ordered}$ \bf({eV/f.u.}) \\
\hline
LCMO ($P2_1/n$)& 0.22&0.0065 \\
LSCMO ($P2_1/n$) & 0.11 & 0.0050 \\
LCMO ($R\overline{3}c$) & 0.29&- \\
LSCMO ($R\overline{3}c$)& 0.18&- \\
\hline
\hline
\end{tabular}
\label{t1}
\end{table*}

\begin{table*}[t]
\caption{The calculated band gap and the corresponding state of metallicity of ordered/disordered LCMO and LSCMO with/without vacancy. For the half-metallic case the band-gap in the minority spin channel has been mentioned.}
\begin{tabular}{| l | c c | c c | }
\hline
\hline
  &  \multicolumn{2}{c|} {\chb{without vacancy}} & \multicolumn{2}{c|}{\chb{with vacancy}}\\
 & State of metallicity &  Band gap & State of metallicity & Band gap \\
 &                      &     eV    &                      &   eV     \\
\hline
Ordered LCMO & Insulator &0.35 & Insulator & 0.518\\
Disrdered LCMO & Half metal &1.38 & Insulator & 1.135\\
Ordered LSCMO & Half metal & 0.80& Insulator & 0.143 \\
Disordered LSCMO & Half metal &0.72 & Insulator & 0.325 \\
\hline
\hline
\end{tabular}
\label{t2}
\end{table*}


\subsubsection{In the presence of vacancy} 
In the following we study the effect of oxygen vacancy on the dominant monoclinic phase of LCMO and LSCMO. We create one vacancy per unit cell of LCMO and LSCMO, which culminates into a vacancy concentration of 8.33\%. The number of configurations to be investigated depends largely on the symmetry of the cell. It is important to check the dynamical stability of these O-vacancy containing LCMO and LSCMO. We have therefore calculated the corresponding phonon density of states for these systems (as presented in the SM\cite{supplemental}), wherein the absence of any imaginary modes indicates the stability of these structures in the presence of defects. We perform total energy calculations on all crystallographically inequivalent configurations and determine the lowest energy structure which defines the crystal structure of the O-site vacancy containing LCMO and LSCMO in their ordered and disordered phases (right panel of FIG. 1). The details of the configurations have been elucidated in the SM\cite{supplemental}. The calculated energy difference between the energetically most favoured disordered structure and the ordered structure of LCMO in the presence of 8.33\% O-site vacancy is found to be 0.0065 eV/$f.u.$ (75 K/$f.u.$) which is negligibly small. This difference is further reduced with Sr doping (0.005 eV/$f.u.$) showing that disorder is highly likely to be there and play a role at room temperature. These results are tabulated in TABLE I. The complete crystal structure data as obtained by first-principles optimization are presented in TABLE S1 of the SM\cite{supplemental}.
\subsection{Electronic structure} 
\subsubsection{In the absence of vacancy}
We perform first-principles calculations considering a collinear ferromagnetic configuration to determine the electronic structure of pure and O-vacancy induced LCMO and LSCMO, the results of which are presented below. Pristine LCMO is found to be a ferromagnetic insulator (refer to FIG. S5 of SM\cite{supplemental} for the total density of states (TDOS)) with a band gap of 0.35 eV. The Co and Mn ions are in their high spin +2 and +4 valence state respectively and the total moment is 6$\mu_B$/$f.u.$. With the introduction of disorder at the Co/Mn sites, the oxidation states of the transition metal ions are also seen to change, in an attempt to accommodate the disorder. Both Co and Mn ions take up +3 oxidation state. The spin polarized total density of states(TDOS) for disordered LCMO(refer to FIG. S5 of SM\cite{supplemental}) shows that it is half-metallic\cite{inst-metal}. We have also checked the effect of U (U$_{eff}$ - 3eV, 4eV, 5eV, 6eV) on the electronic structure of disordered LCMO. Since both Co and Mn are 3d transition metal ions, any small difference in U between the two ions is not expected to bring in any difference in the qualitative nature of our results\cite{hubbard_U1,hubbard_U2,hubbard_U3}. Further details on the treatment of U may be found in the SM\cite{supplemental}. The oxidation state of the transition metal ions as well as the obtained half-metallicity seem to be robust against the variation of U in the said range\cite{hubbard_U2}. Experimentally the magnetic moment of LCMO is 5.78$\mu_B$/$f.u.$ at 5K and 50 kOe \cite{21} which is close to our calculated value. Calculated total moment and ionic moment per $f.u.$ of ordered and disordered LCMO is presented in TABLE III.
\begin{figure*}[htp]
\centering
\includegraphics[scale=.60]{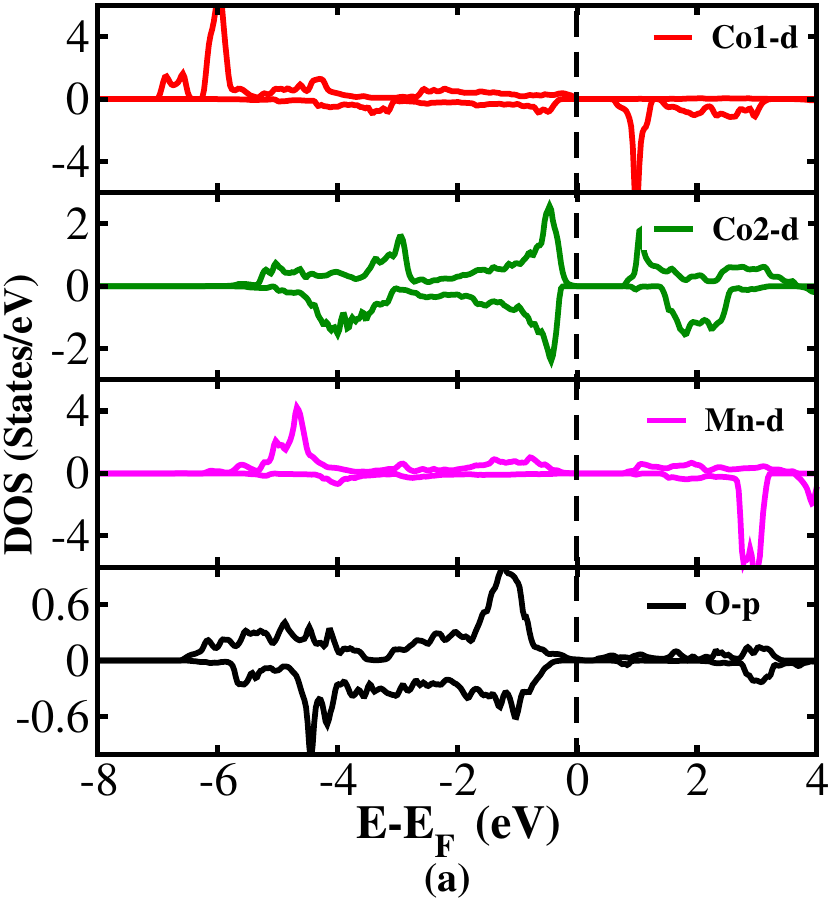}  
\includegraphics[scale=.60]{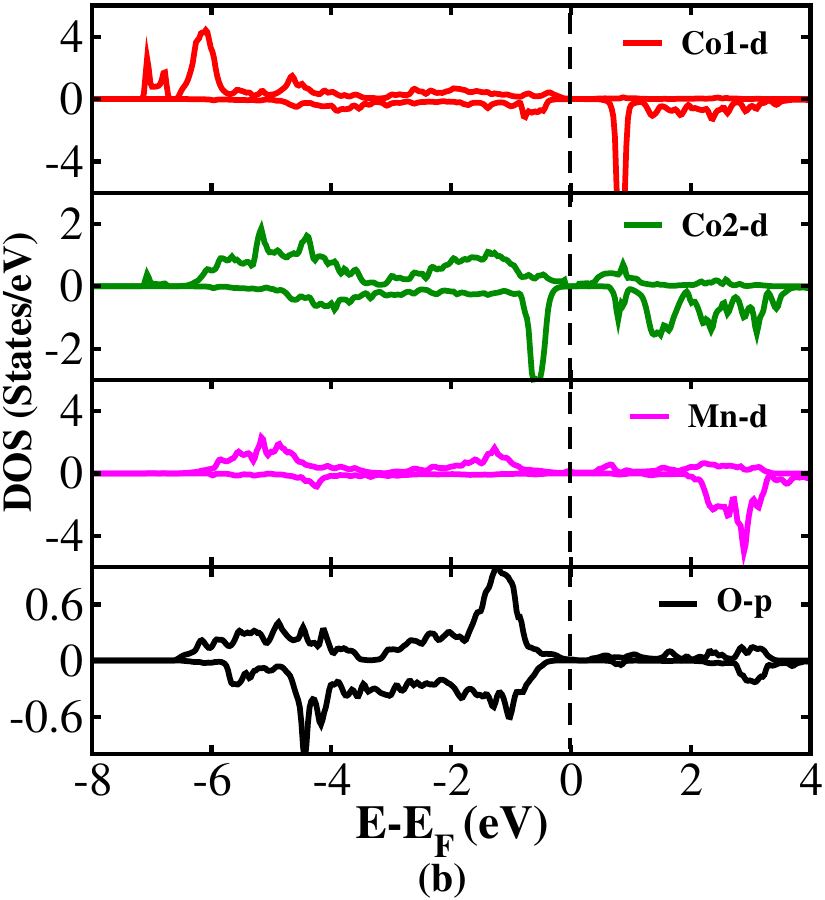}
\caption{Spin polarised partial density of states of Co1-$d$, Co2-$d$, Mn-$d$ and O-$p$ for (a) ordered LSCMO and (b) disordered LSCMO.}
\end{figure*}

\begin{figure*}[htp]
\centering
\includegraphics[scale=.60]{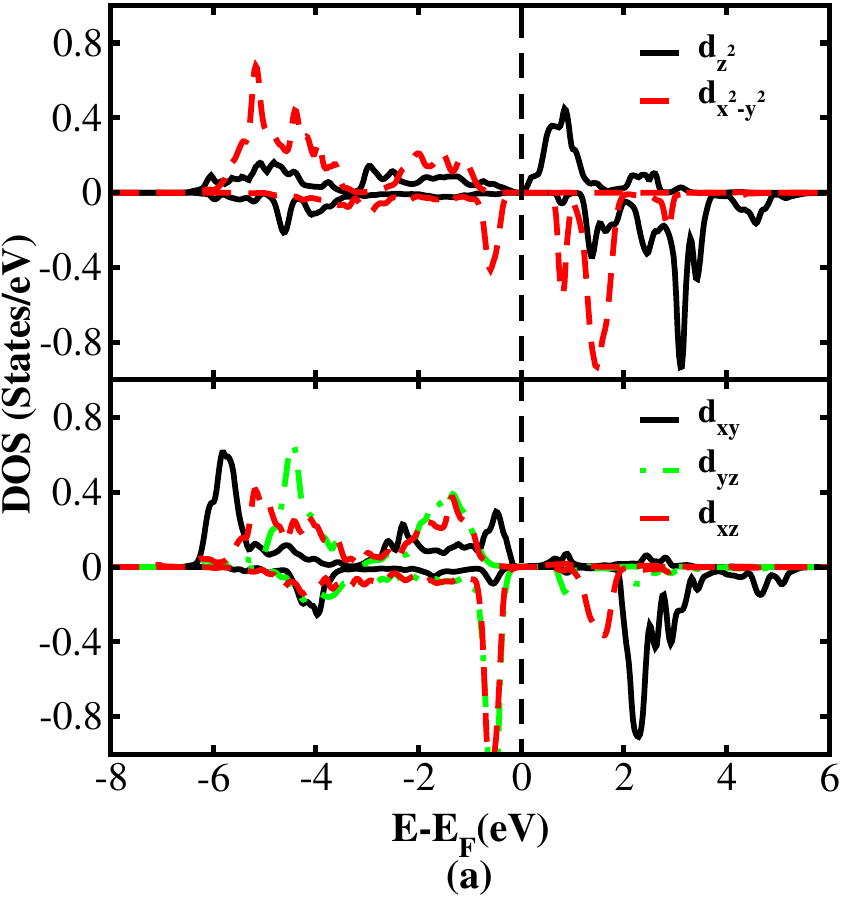}
\includegraphics[scale=.60]{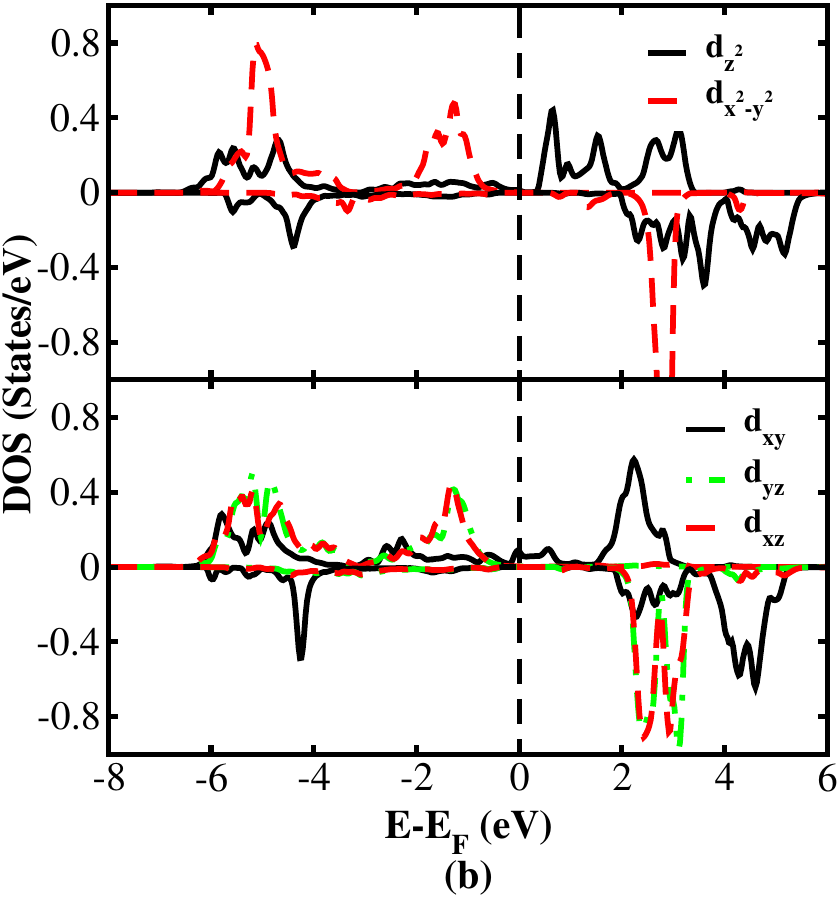}
\caption{Spin polarised partial density of states of (a) Co2-e$_g$ levels (upper panel) and Co2-t$_{2g}$ levels (lower panel), (b) Mn-e$_g$ levels (upper panel) and Mn-t$_{2g}$ levels (lower panel), of disordered LSCMO.}
\end{figure*}

\begin{figure*}[htp]
\centering
\includegraphics[scale=.8]{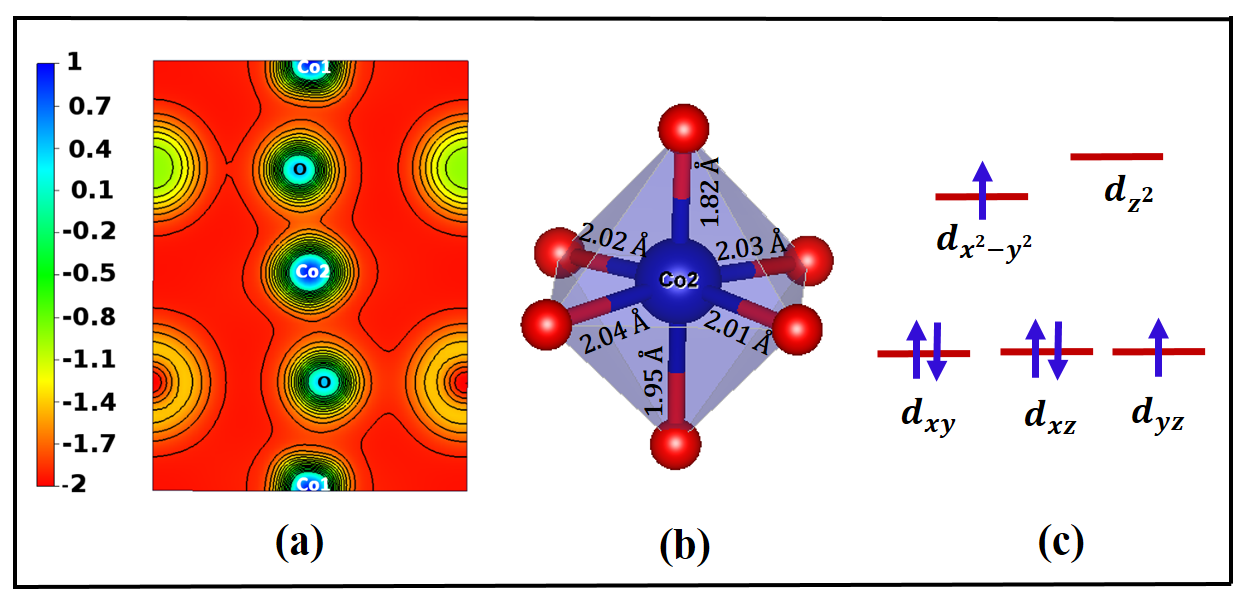}  
\caption{(a) Charge density plot of disordered LSCMO projected onto the \textit{bc} plane. (b) Schematic diagram showing the Co2-O bond lengths in disordered LSCMO. (c) Schematics showing the energy level splitting of Co2-$d$ orbitals and their respective occupancies.}
\end{figure*}
We next consider Sr doped LCMO. Replacement of two La$^{3+}$ by two Sr$^{2+}$ in the system creates charge imbalance, which is expected to be neutralized by the readjustment of the charge state of transition metal cations (Co, Mn). Spin-polarized TDOS of ordered LSCMO (refer to S6 of SM\cite{supplemental}) shows it to be a half-metal. Analysis of the partial density of states(PDOS) (FIG. 2(a)) of Co and Mn shows: (i) there is very strong hybridization between Co/Mn-$d$ with O-$p$, (ii) both the Mn ions are in high-spin +4 state, (iii) Co is in mixed valence state, with Co1 in high-spin +2 charge state (magnetic moment = 2.8$\mu_B$) and Co2 in low-spin +3 charge state (magnetic moment = 0.019$\mu_B$). The resultant charge state for each cation however does not comply with a charge-neutral system. This is because the material we are considering is highly covalent, and the extra hole that is apparently unaccounted for, is taken care of simultaneously by the Co$^{2+}$-like ion and the oxygen atoms surrounding it. This is evident from the increased magnetic moment at the O-atoms connected to Co$^{2+}$ by about 0.15$\mu_B$. Therefore, charge imbalance caused by hole doping is primarily taken care by the Co ions. A close inspection of the crystal structure helps to unravel the reason behind Co2 taking up a low spin state. The average bond length of Co2-O along \textit{x-y} plane is found to be shorter than the Co1-O bond by 0.079\r{A} and along the z-direction by 0.11\r{A} suggesting that the Co2O$_6$ octahedra is much smaller as compared to Co1O$_6$ octahedra. This leads to greater octahedral field splitting, such that the energy difference between the t$_{2g}$ and e$_{g}$ levels is greater than the energy cost of putting two electrons at the same site. As a result electrons prefer to accommodate themselves in t$_{2g}$ level. We next consider disordered LSCMO. Spin-polarized TDOS of disordered LSCMO also shows it to be half-metallic (refer to S8 of SM\cite{supplemental} for the band structure). Analysis of PDOS(FIG. 2(b) and FIG. 3) of Co and Mn suggests: (i) there is strong hybridization between Co/Mn-$d$ ions with O-$p$; (ii) the charge and spin state of Mn ions remain unaltered, i.e, +4 high spin state; (iii) charge and spin state Co ions are altered, Co1 is found high spin +2 charge state (magnetic moment = 2.87$\mu_B$) and Co2 is found in intermediate spin +3 charge state (magnetic moment=1.97$\mu_B$). It is to be noted that because of very strong co-valency between O-$p$ and Co/Mn-$d$, the calculated magnetic moments at the transition metal sites do not agree very well with that calculated using an ionic model. Analysis of the crystal structure shows that one of the Co2-O bonds along the z axis is much smaller than the other (FIG. 4(b)). As a result the covalency is very strong along this shortened bond and the hole at the Co site is shared to a great extent by the O ion of this bond. This is clearly evident from the charge density plot in FIG. 4(a). We find that the magnetic moment on this O-atom is higher than the rest by 0.3$\mu_B$. In order to unravel the microscopic origin of the spin state of Co2, we perform further analysis of the crystal structure of disordered LSCMO. We find that the average bond length along the \textit{x-y} plane is larger that along the \textit{z} direction (FIG. 4(b)). This makes the energy of $d_z^2$ level rise in comparison to $d_{{x^2} {-}{y^2}}$, leading to an intermediate spin state. The schematic energy level diagram is shown in FIG. 4(c). The state of metallicity and the value of the PBE band-gap(wherever applicable) is presented in TABLE II. The moments at Co/Mn sites and the corresponding total moment/f.u. considering a ferromagnetic state is summarized in TABLE III. 
\begin{table*}[t]
\caption{The charge and spin state of the transition metal (TM) cations, the calculated magnetic moment at each TM site, and the magnetisation per formula unit (f.u.) of ordered/disordered LCMO and LSCMO with/without vacancy. All calculations have been performed presuming a ferromagnetic state.}
\begin{tabular}{| l | c c c c | c c c c |}
\hline
\hline
 & && \bf{\chb{without vacancy}} &&&& \bf{\chb{with vacancy}}&\\
 & Ionic element&Spin& Total moment & Ionic moment&Ionic element &Spin&Total moment&Ionic moment \\
  & &state&\bf{$\mu_B$/f.u.}&\bf{{$\mu_B$}}&&state&\bf{$\mu_B$/f.u.}&\bf{{$\mu_B$}}\\
\hline
Ordered LCMO &&& 6.00 & &&&7.00 &\\
           &Mn$^{4+}$ &HS&& 2.95&Mn$^{3+}$ &HS&& 3.89 \\
           &Co$^{2+}$ &HS&& 2.80&Co$^{2+}$ &HS&& 2.65 \\
Disrdered LCMO &&& 5.00 &&&&7.00& \\
             &Mn$^{3+}$ &HS&& 3.33&Mn$^{3+}$ &HS&& 3.88 \\
             &Co$^{3+}$ &LS&& 0.23&Co$^{2+}$ &HS&& 2.61 \\
Ordered LSCMO &&& 5.00&&&& 6.00& \\
            &Mn$^{4+}$ &HS&& 2.90&Mn$^{3+}$&HS&&3.28 \\
            &Co1(Co$^{2+}$) &HS&& 2.80&Co$^{2+}$ &HS&& 2.54 \\
            &Co2(Co$^{3+}$) &LS&& 0.019&&&& \\
Disordered LSCMO &&&6.00&&&& 7.00&\\
               &Mn$^{4+}$&HS&&2.96&Mn$^{3+}$&HS&&3.76\\
               &Co1(Co$^{2+}$)&HS&&2.87&Co$^{2+}$ &HS&& 2.95\\
               &Co2(Co$^{3+}$)&IS&&1.97&&&&\\
\hline
\hline
\end{tabular}
\label{t3}
\end{table*}

\subsubsection{In the presence of vacancy} 
\begin{figure*}[htp]
\centering
\includegraphics[scale=.60]{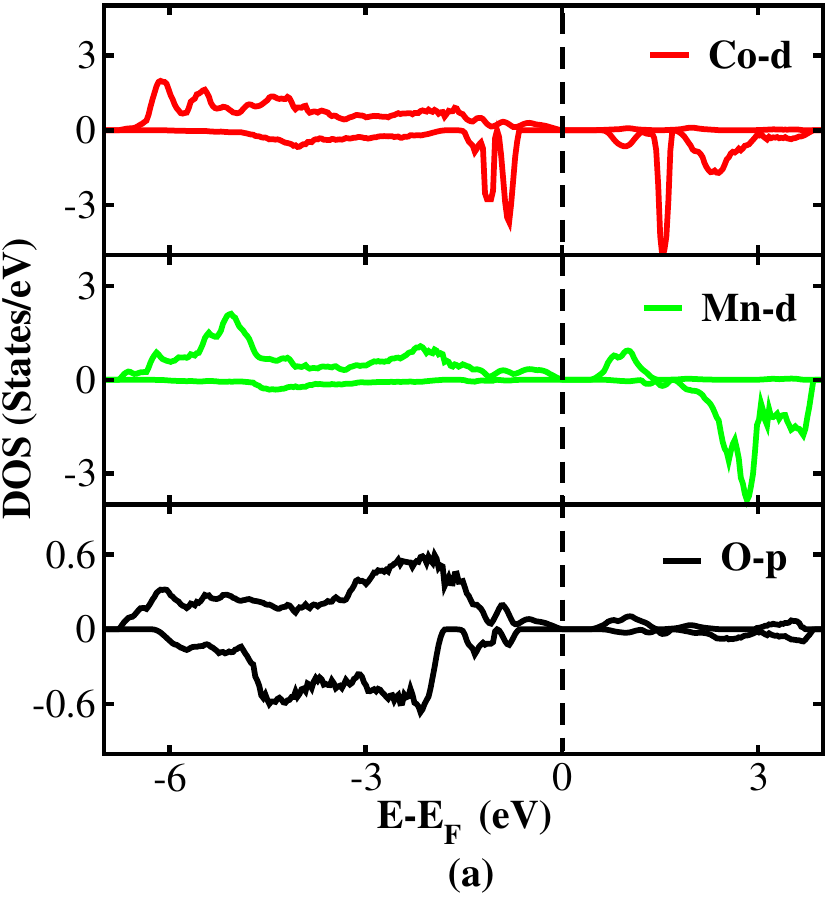}  
\includegraphics[scale=.60]{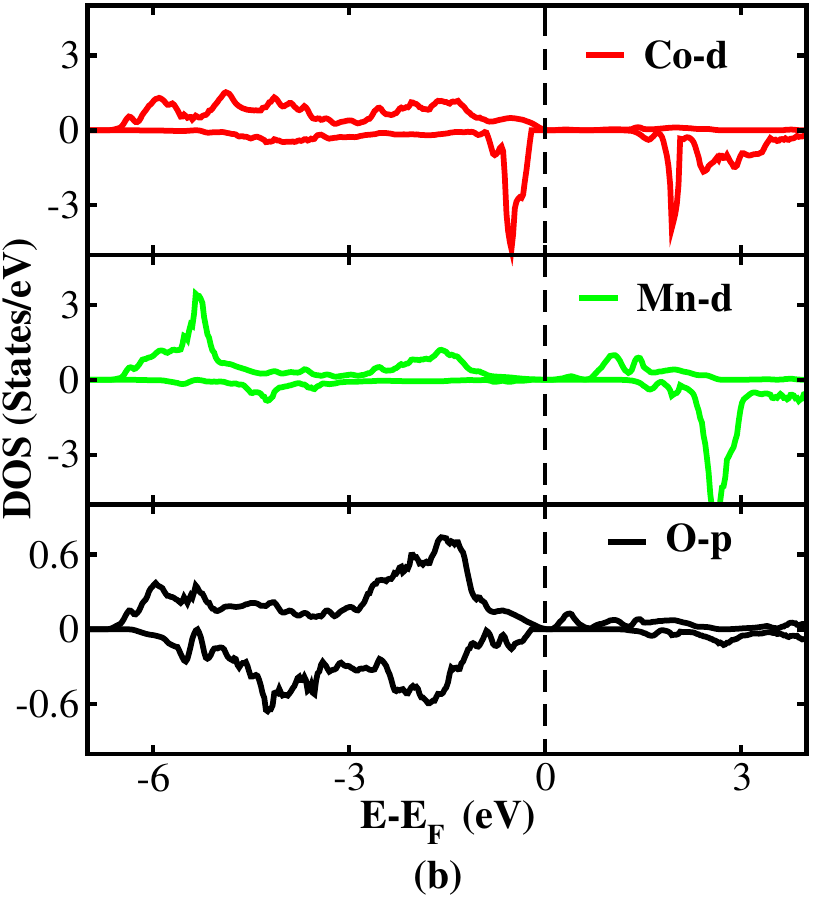}
\caption{Spin polarised partial density of states of Co-$d$, Mn-$d$ and O-$p$ for (a) ordered LCMO and (b) disordered LCMO with vacancy.}
\end{figure*}
 In order to delineate the effect of O-site vacancy on the properties of LCMO and LSCMO, we have created one vacancy per unit cell which culminates into 8.33\% vacancy concentration at O-site. Our calculations show that with the introduction of O-site vacancy, appropriate number of Mn$^{4+}$ converts to Mn$^{3+}$ to maintain the overall charge neutrality. The calculated spin-polarized TDOS of ordered La$_2$CoMnO$_{5.2}$ is shown in the SM\cite{supplemental}. We find that even after the introduction of defects into the system, it retains its ferromagnetic insulating state, the calculated band gap being 0.518 eV (TABLE II) which is larger than without vacancy case. The PDOS of Co and Mn ions(FIG. 5) show that they appear in their +2 and +3 charge state respectively. The calculated magnetic moments at Co (2.65$\mu_B$/f.u.) and Mn (3.89$\mu_B$/f.u.) site also ascertain the same. The total magnetic moment of the system comes out to 7$\mu_B$/f.u. (refer to TABLE III). Electronic structure calculation of the lowest energy O-site vacancy containing disordered LCMO structure shows that introduction of vacancy opens up a gap at the fermi level converting the otherwise half-metallic system into an insulator with a band gap of 1.135 eV. Co and Mn ions with an magnetic moment of 2.61$\mu_B$/f.u. and 3.88$\mu_B$/f.u. are found in its +2 and +3 charge state respectively. On doping 50\% Sr at La sites in the O-site vacancy containing structure, the electronic structure calculation gives an insulating solution (refer to S6 of SM\cite{supplemental} for TDOS) with a band gap of 0.143 eV. The Co and Mn ions with an average moment of 2.54$\mu_B$/f.u. and 3.28$\mu_B$/f.u. is found in +2 and +3 charge state respectively. Unlike its vacancy-free counterpart, the introduction of Sr$^{2+}$ in place of La$^{3+}$ in the vacancy containing system does not seem to alter the charge state of Co. The hole that is introduced as a result of Sr doping at adjacent La sites is adjusted by both the oxygen atoms and the Co ions, such that there is no significant change in the charge state of Co and it effectively remains in the +2-like charge-state. The magnetic moment at the nearby oxygen atoms is enhanced by 0.26$\mu_B$/f.u. and the Co ions is enhanced by 0.35 $\mu_B$/f.u., which is a testimony to the strong covalency that prevails in these double perovskite oxides. Similarly when we study the electronic structure of anti-site disorder containing LaSrCoMnO$_{5.2}$, we find that introduction of vacancy introduces a band gap of 0.325 eV and the PDOS character as well as the magnetic moments of Co (2.95$\mu_B$/f.u.) and Mn (3.76$\mu_B$/f.u.) ions suggest that they are in +2 and +3 valence state with the extra hole coming from Sr doping adjusted in both Co and O ions. The state of metallicity and the value of the PBE band-gap (wherever applicable) is presented in TABLE II. The moments at Co/Mn sites and the corresponding total moment/f.u. considering a ferromagnetic state is summarized in TABLE III. 
 
\subsection{Magnetic structure and Curie temperature} 
In this section we discuss our results on the evolution of the magnetic structure and corresponding transition temperature (T$_c$) of ordered and disordered LSCMO in the presence of Sr-doping and/or O-site vacancy. Introduction of defect generally has a significant impact on the bond-angles and bond-lengths of the system - they alter to accommodate the defect. This is expected to influence the magnetic structure of the system as well. 
The O-site vacancy gives rise to electron doping and replacing La$^{3+}$ by Sr$^{2+}$ leads to hole doping. In order to decipher the nature of magnetism in the presence of doped electrons and holes, we have calculated the energetics of all possible collinear magnetic configurations in an unit cell (refer to TABLE IV) of the system with and without doping. The calculated energies of each spin configuration are then mapped onto an Ising Hamiltonian\cite{Ising}: 
\begin{equation}
     H=-\sum_{ij} J_{ij}\vec{S}_i\cdot \vec{S}_j
\end{equation}
\noindent where J$_{ij}$ is the magnetic exchange interaction between spins at i$^{th}$ and j$^{th}$ sites, having effective spin values $S_i$ and $S_j$ respectively. We have primarily considered the nearest neighbour interactions which are expected to be the dominant ones, namely, J$_1$: interaction between two nearest neighbour Co1 and Co2 ions, J$_2$: interaction between two nearest neighbour Mn1 and Mn2 ions, J$_3$ being the interaction between two nearest neighbour Co1 and Mn1 ions; J$_4$, the interaction between two nearest neighbour Co1 and Mn2 ions, J$_5$: interaction between two nearest neighbour Co2 and Mn1 ions, and J$_6$, the interaction between two nearest neighbour Co2 and Mn2 ions. The corresponding magnetic interactions are schematically shown in FIG. 6.
\begin{figure}
\begin{center}
\rotatebox{0}{\includegraphics[width=0.4\textwidth]{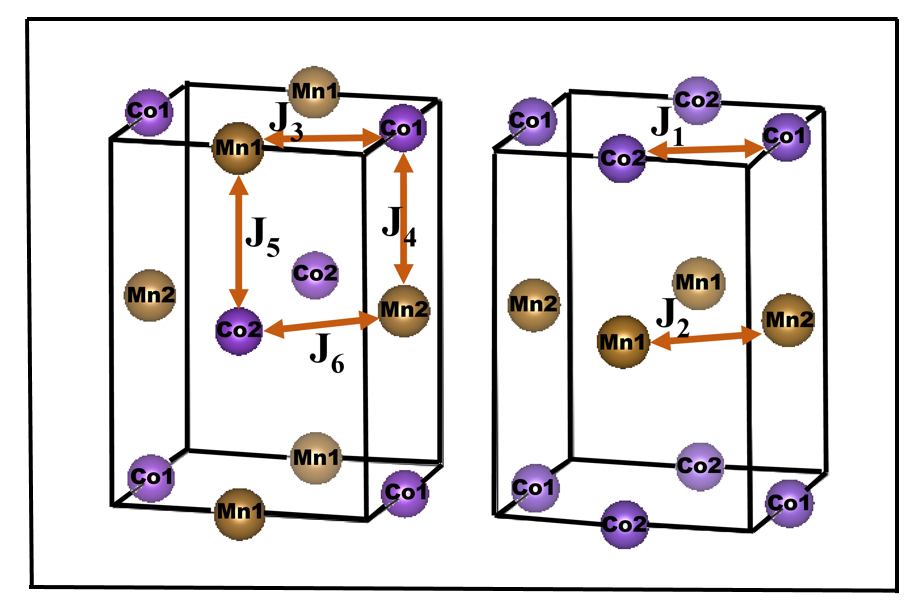}}
\end{center}
\caption{Magnetic exchange interactions between different transition metal ions in disordered LCMO.} 
\end{figure}
\subsubsection{In the absence of vacancy}
In agreement with experiments \cite{21}, the ferromagnetic spin arrangement is found to be the ground state magnetic configuration for ordered LCMO. This would give rise to in-plane and out of plane ferromagnetic coupling between the nearest neighbour Mn$^{4+}$ and Co$^{2+}$. The values of the interaction parameters have been tabulated in TABLE V. J$_1$ and J$_2$ interactions are irrelevant here, as, in the ordered structure there are no nearest neighbour Co-Co or Mn-Mn interactions. Out of all the interactions considered, in plane coupling J$_3$ turns out to be the largest with a numerical value of 3.53 meV. For disordered LCMO, since both the Co$^{3+}$ ions go into a low spin state resulting in zero moment at the Co sites, the only interaction that is active is J$_2$, which represents the Mn1-Mn2 exchange interaction. The calculated value of J$_2$ is 1.76 meV. With the introduction of Sr in the system, Co1(Co$^{3+}$) is found in low spin state. As a result interaction parameters (J$_1$, J$_3$, J$_4$) involving Co1 are inactive. The remnant interactions are found to be ferromagnetic. However, when we consider disorder at Co/Mn sites in LSCMO, two of the interactions J$_1$, J$_2$ are antiferromagnetic, whereas, J$_4$, J$_5$ are ferromagnetic. It is seen that the value of exchange interactions tabulated in TABLE V are larger for Sr-doped system.

The corresponding transition temperatures (T$_C$) (scaled with respect to ordered LCMO) obtained using mean-field approximations are listed in TABLE V. Although mean-field calculations overestimate the Curie temperatures, we find that it provides an accurate trend nonetheless, with the scaled values close to the experimental ones. In fact the approximation is not too bad as the spin values are larger (S $>$ 1/2) and the system is 3D. The details of the theoretical backdrop based on which we have performed our mean-field calculation is presented in the SM\cite{supplemental}.  

\subsubsection{In the presence of vacancy}
We have also considered the effect of O-site vacancy on the magnetic interaction parameters. The Mn and Co ions are found in +3 and +2 high spin state respectively in O-site vacancy containing LCMO and LSCMO (refer to TABLE III). Our calculated exchange interactions, as presented in TABLE V, show that both their magnitude and sign are affected by electron doping. The magnetic transition temperatures (T$_C$), obtained using mean field method (details can be found in SI), are tabulated in the TABLE V. The T$_C$-s are found to get reduced as a result of electron doping.    

\begin{table}
\caption{All possible magnetic configurations per unit cell.}
\begin{tabular}{|c c c c c|}
\hline
\hline
 &Co1&Co2&Mn1&Mn2\\
\hline
FM&$\uparrow$&$\uparrow$&$\uparrow$&$\uparrow$ \\
config.1&$\downarrow$&$\uparrow$&$\uparrow$&$\uparrow$ \\
config.2&$\uparrow$&$\downarrow$&$\uparrow$&$\uparrow$ \\
config.3&$\uparrow$&$\uparrow$&$\downarrow$&$\uparrow$ \\
config.4&$\uparrow$&$\uparrow$&$\uparrow$&$\downarrow$ \\
config.5&$\uparrow$&$\downarrow$&$\downarrow$&$\uparrow$ \\
config.6&$\uparrow$&$\uparrow$&$\downarrow$&$\downarrow$ \\
config.7&$\uparrow$&$\downarrow$&$\uparrow$&$\downarrow$ \\
\hline
\hline
\end{tabular}
\label{t4}
\end{table} 
\begin{table*}[t]
\caption{Magnetic interactions between nearest neighbour atoms in the unit of meV. The Curie temperature(T$_C$) for the different structures have been scaled with respect to the T$_C$ of ordered LCMO.}
\begin{tabular}{| l c | c c c c c c | c c c | }
\hline
\hline
&&&&&&&&&\multicolumn{2}{c|}{ Scaled Curie Temperature}\\
\hline
&&J$_1$&J$_2$&J$_3$&J$_4$&J$_5$&J$_6$&&Mean field&Experiment\\
\hline
\bf{\chb{Without vacancy}}&&&&&&&&&&\\
\hline
Ordered LCMO && &  & 3.53 & 1.80 & 2.86 & 1.43 && 1 & 1\cite{25} \\
disordered LCMO && & 1.76 & & & & && 0.77 & 0.63\cite{25} \\
Ordered LSCMO && & & & & 8.94 & 2.22 && 0.92 & $\approx$ 0.91\cite{25} \\
disordered LSCMO && -10.22 & -2.16 & & 8.91 & 4.47 &  && 0.88&\\
\hline
\bf{\chb{With Vacancy}}&&&&&&&&&&\\
\hline
Ordered LCMO && & & -1.64 & -5.87 & 4.66 & 3.04 && 0.39& \\
disordered LCMO && -3.25 & 1.29 & & 2.76 & -1.09 & && 0.41& \\
Ordered LSCMO &&  & & 3.21 & -0.69 & 2.04 & -1.063 && 0.53& \\
disordered LSCMO && 3.68 & -1.60 & & 3.63 & -3.12 & && 0.46& \\
\hline
\hline
\end{tabular}
\label{t5}
\end{table*} 
\section{Linear optical Properties}

In order to study the effect of electron and hole doping on the linear optical properties of LCMO, we calculate the complex dielectric function $\epsilon(\omega)$ of the material using first-principles DFT. The optical properties of a solid state material is a manifestation of its dynamical interaction with incident electromagnetic wave and hence can be studied using the time-dependent perturbation theory. More details about the theory can be found in the SM\cite{supplemental}.

Since the monoclinic space group, namely, P2$_{1}$/n is dominant in LCMO and LSCMO, in the following we have investigated the optical properties of these materials only in their monoclinic symmetry. The monoclinic distortion ($\beta$) in LCMO is very small, i.e, its deviation from $90^{\circ}$ is not significant (refer to TABLE S1 of the SM\cite{supplemental}). As a result, the off-diagonal elements of the dielectric function are very small compared to the diagonal elements\cite{dielectric7}. Therefore, in the following analysis of the linear optical properties, we ignore the off-diagonal elements and consider only the major components, namely, {\bf $\epsilon_{zz}$} and {\bf $\epsilon_{xx}$} (which, respectively, correspond to an electric field {\bf E} parallel and perpendicular to the {\bf {\textit C}} axis) of the dielectric function. The real and imaginary part of these are plotted as a function of energy for further analysis. Thereafter, the static dielectric constants Re$\epsilon(0)$ are determined which are employed to calculate the uniaxial anisotropy, defined as [$\delta\epsilon=(\epsilon_{zz}-\epsilon_{xx})/\epsilon_{tot}$]

We have also determined the birefringence which is a very important parameter as far as optical properties are concerned and is defined as {\bf $\Delta n(\omega)$= $n_e(\omega)$-$n_o(\omega)$}, where n$_e$ and n$_o$ are the extraordinary and ordinary refractive indices respectively.

\subsubsection{In the absence of vacancy}
The $\epsilon_{xx}$ and $\epsilon_{zz}$ components of the real and imaginary part of the dilectric function for ordered and disordered LCMO and LSCMO are shown in FIG. 7 (a-d). For ordered LCMO, (i) the upper panel of FIG. 7(a) and the inset therein shows that the anisotropy in the real part of the dielectric function, though present, is insignificant; (ii) the fundamental peak in the imaginary part for both $\epsilon_{xx}$ and $\epsilon_{zz}$ appear at 0.62 eV; (iii) the calculated uniaxial anisotropy as shown in TABLE VI is therefore very small. Whereas for disordered LCMO, (i) the mis-match between the real parts of $\epsilon_{xx}$ and $\epsilon_{zz}$ is found to increase; (ii) The fundamental peak for $\epsilon_{xx}$ appears at 0.09 eV and for $\epsilon_{zz}$, it appears at 0.07 eV; (iii) The increase in anisotropy is also evident in the calculated values of uniaxial anisotropy as presented in TABLE VI. With the introduction of Sr in the system, the optical anisotropy, determined from the mis-match of the $\epsilon_{xx}$ and $\epsilon_{zz}$ components of the dielectric function (FIG. 7(c) and 7(d)), is found to increase in comparison to its undoped counterpart, i.e, pure LCMO. The calculated uniaxial anisotropy obtained from the real part of the static dielectric function (refer to TABLE VI) shows a significant enhancement with Sr doping for both ordered and disordered LCMO samples. The calculated values of the birefringence for all the above-mentioned samples are presented in TABLE VI.  
 
\begin{figure*}
\centering
\onecolumngrid
$\begin{array}{cccccccc}
\includegraphics[scale=.30]{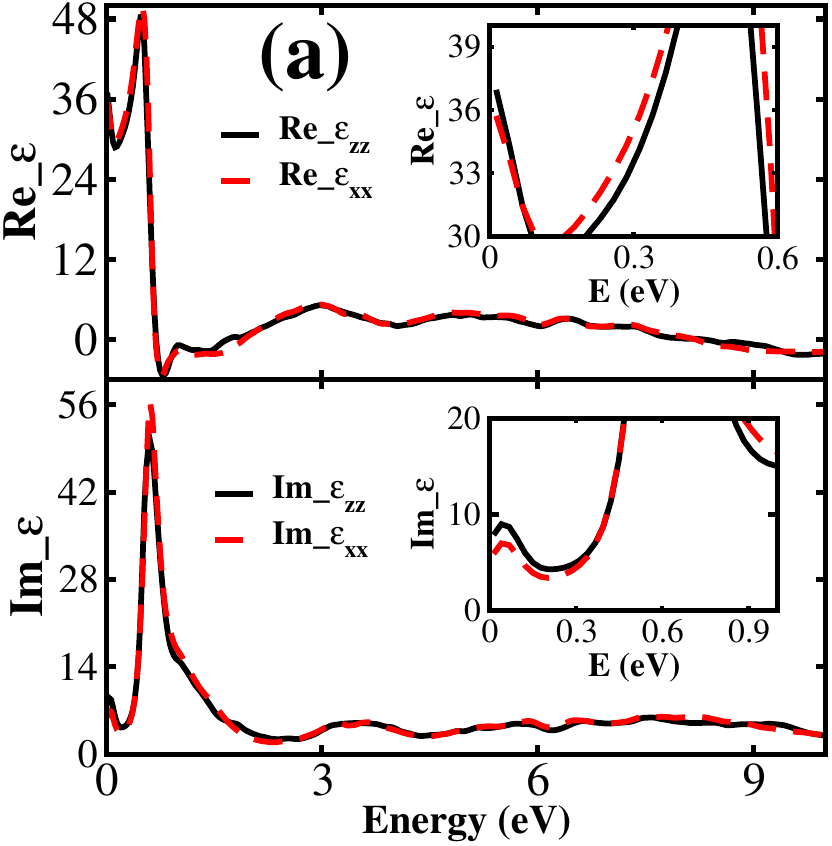} & 
\includegraphics[scale=.30]{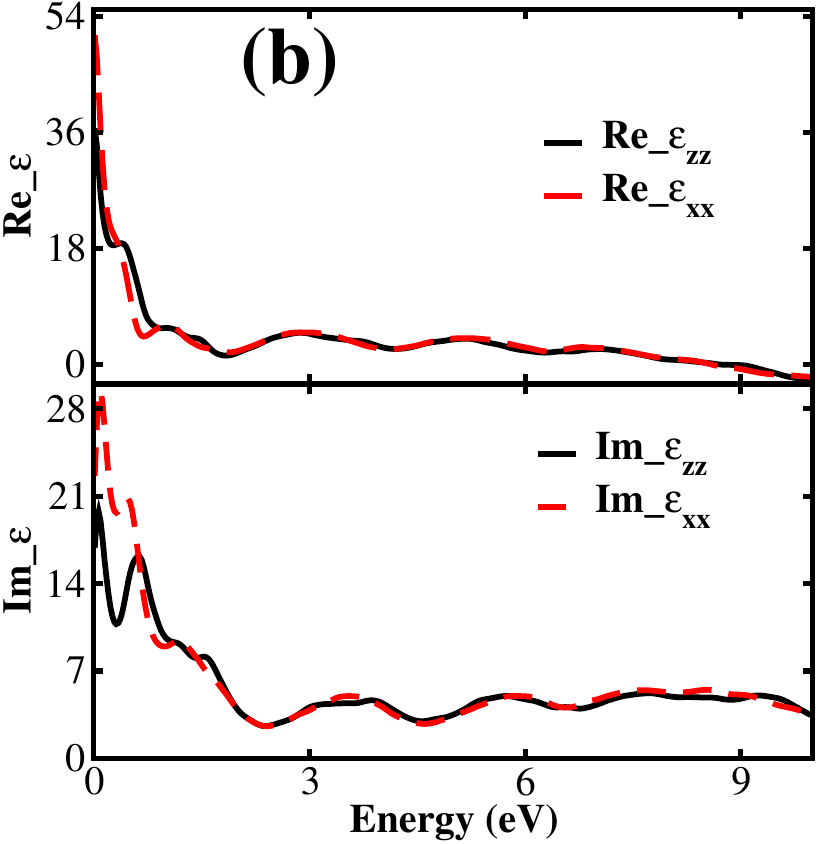} &
\includegraphics[scale=.30]{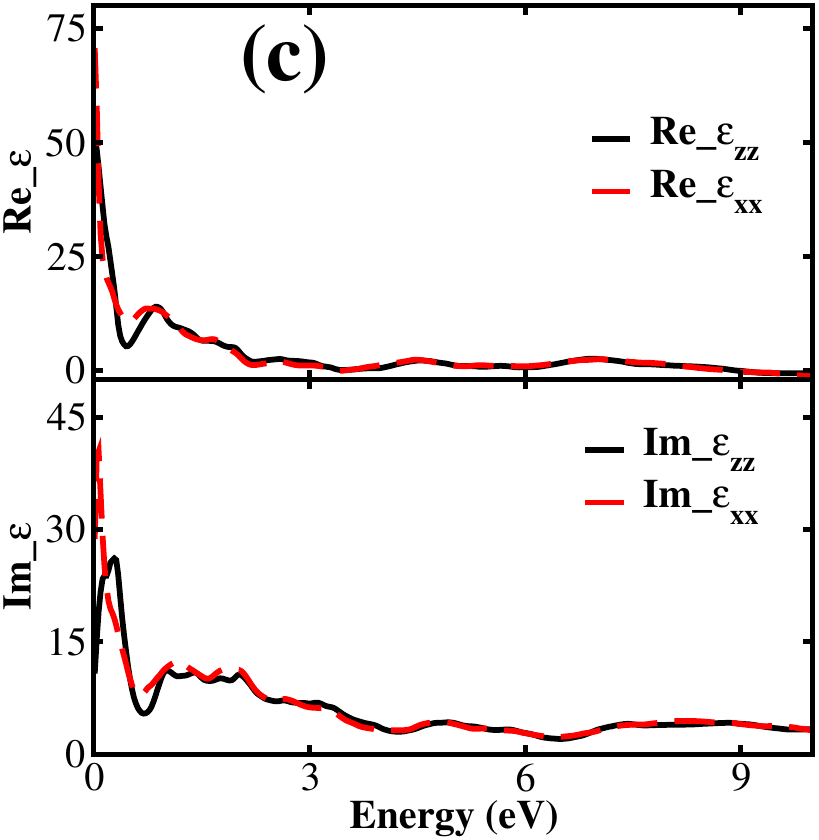}  &
\includegraphics[scale=.30]{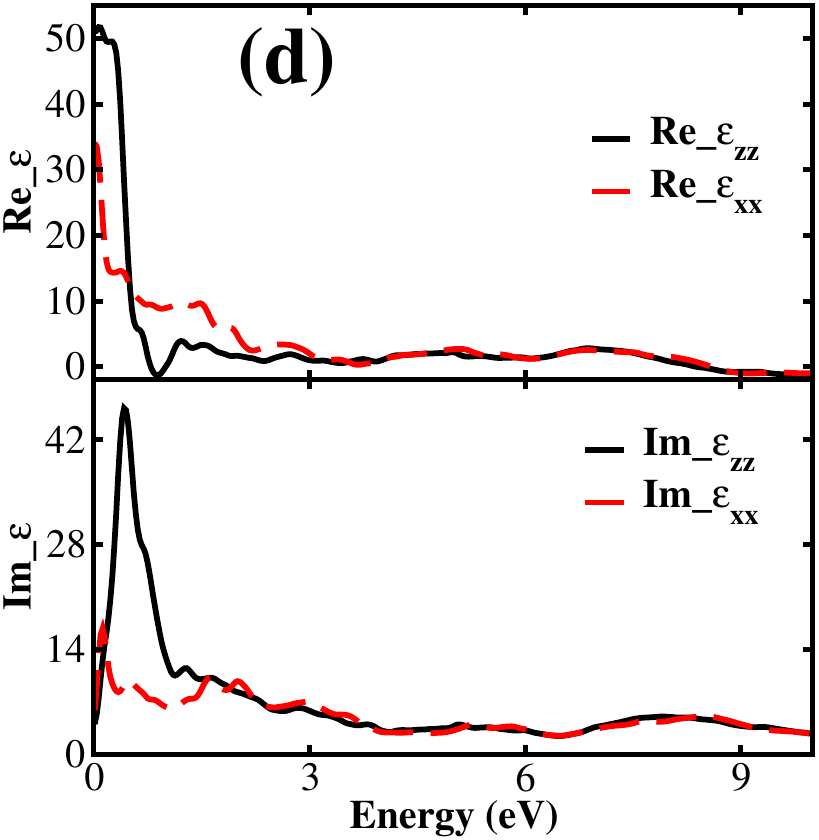} \\
\includegraphics[scale=.30]{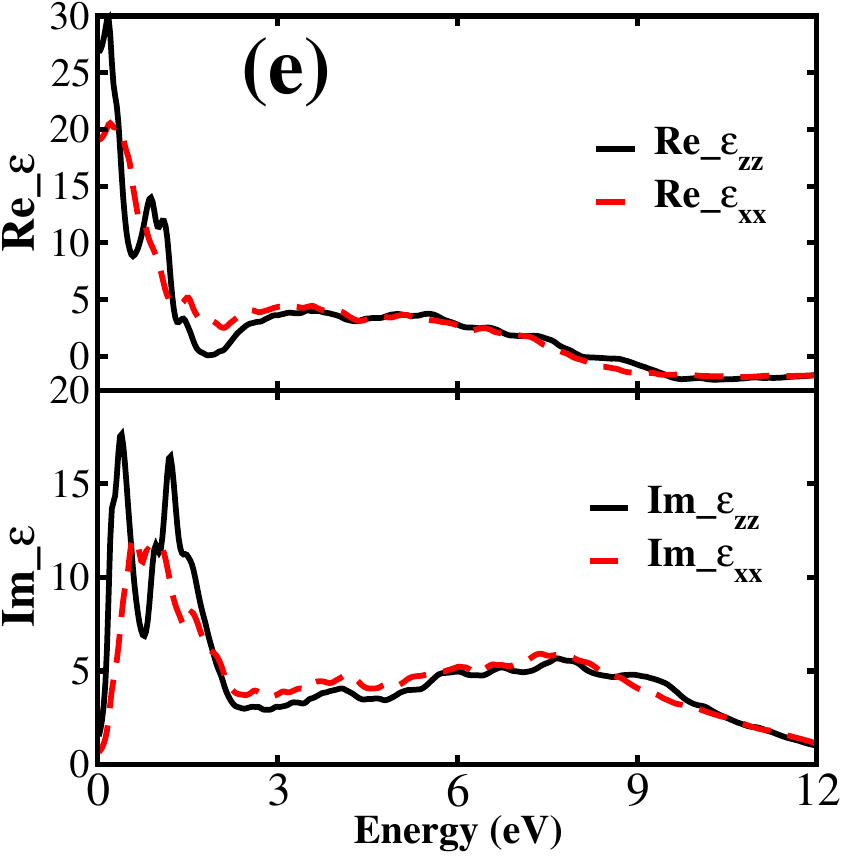} & 
\includegraphics[scale=.30]{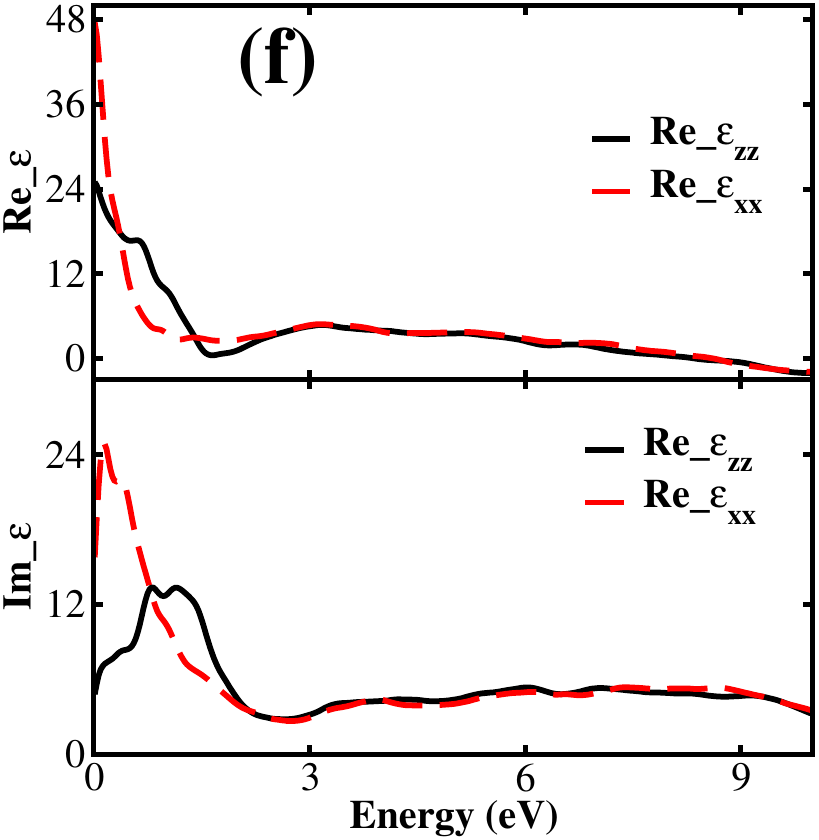} &
\includegraphics[scale=.30]{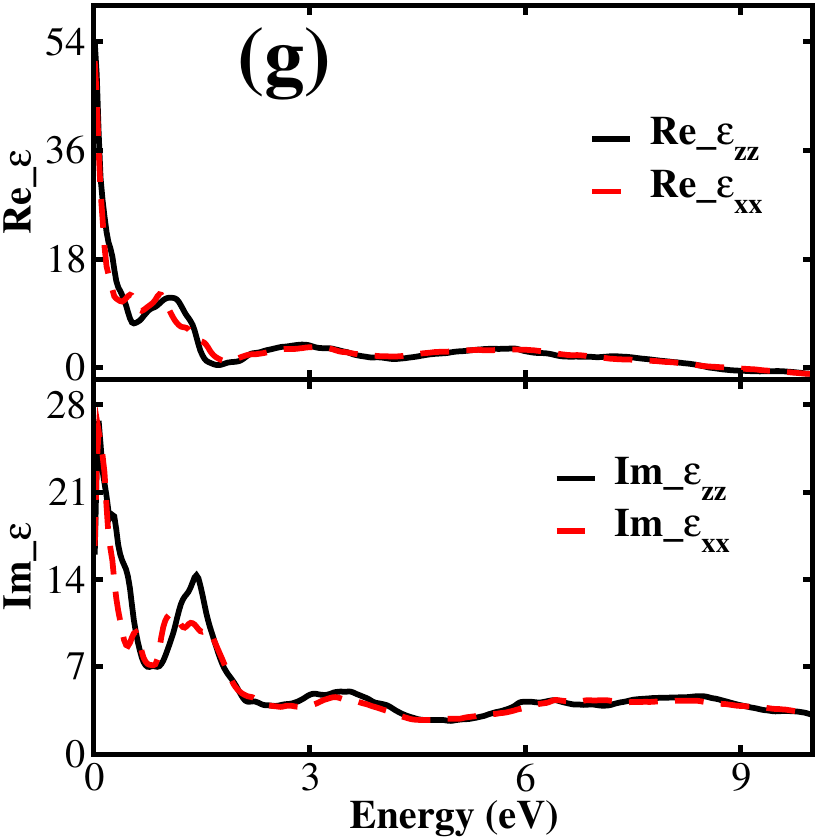}  &
\includegraphics[scale=.30]{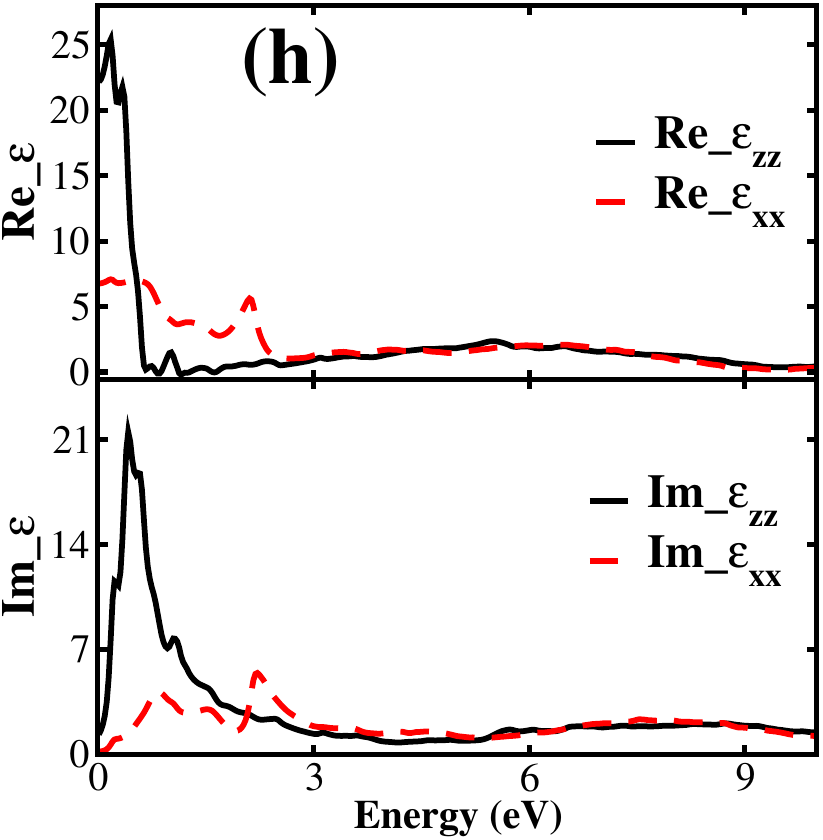} \\
\end{array}$
\caption{Real and Imaginary parts of the dielectric function of (a) ordered LCMO (b) disordered LCMO (c) ordered LSCMO and (d) disordered LSCMO (e) ordered LCMO with O-site vacancy (f) disordered LCMO with O-site vacancy (g) ordered LSCMO with O-site vacancy and (h) disordered LSCMO with O-site vacancy. }
\twocolumngrid
\end{figure*}
 
\subsubsection{In the presence of vacancy}
The real and imaginary parts of dielectric function for ordered and disordered LCMO in the presence of 0.8\% O-site vacancy is shown in FIG. 7(e-h). For the imaginary part of the dielectric function, the fundamental peak of ordered LCMO with O-site vacancy appears at 0.65 eV and 0.41eV for $\epsilon_{xx}$ and $\epsilon_{zz}$, respectively (refer to lower panel of FIG. 7(e)). The presence of vacancy, therefore, introduces an anisotropy in the optical properties of the system, which is otherwise very close to being isotropic for all practical purposes in its purest form, as discussed previously. The calculated value of uniaxial anisotropy is 0.164, an order of magnitude higher than that of the pure, vacancy-free structure. For the disordered LCMO containing O-site vacancy, the fundamental peak appears at 1.96 eV for $\epsilon_{xx}$ and at 0.82 eV for $\epsilon_{zz}$. The uniaxial anisotropy, as presented in TABLE VI, shows that there is a further increase in anisotropy with the introduction of disorder at the Co and Mn sites in the presence of O-site vacancy. However, when Sr is doped into the ordered structure containing O-site vacancy, the optical anisotropy brought in by the electron doping (due to O-site vacancy) is found to get cancelled by the introduction of holes (through doping Sr$^{2+}$ at La$^{3+}$ sites) quite surprisingly. The calculated value of uniaxial anisotropy is quite low(0.03) and is also reflected in the plots of $\epsilon_{xx}$ and $\epsilon_{zz}$ vs energy, presented in FIG. 7(g). When the imaginary part of the dielectric function is plotted as a function of energy, the fundamental peak appears at 0.073 eV for both the components, namely, $\epsilon_{xx}$ and $\epsilon_{zz}$. However, when we consider anti-site disorder at the Co and Mn sites, in addition to  Sr-doping at La-site and vacancy at O-site, the system becomes highly anisotropic as reflected by the significantly high value of uniaxial anisotropy(0.541). The fundamental peaks in the plot of the imaginary component of the dielectric function for $\epsilon_{xx}$ and $\epsilon_{zz}$ appear at 0.84 eV and 0.43 eV respectively, as seen in the lower panel of FIG. 7(h). The refractive indices of these samples have also been calculated and the results show that there is considerable anisotropy in the low energy regime. The calculated birefringence values are presented in TABLE VI.
\begin{table}
\caption{The uniaxial anisotropy and birefringence value for LCMO and LSCMO.}
\begin{tabular}{| l | c c c |}
\hline
\hline
  & Uniaxial anisotropy & Birefringence&  \\
 &value $\delta\epsilon$& value $\Delta n(\omega)$&\\
 &$=(\epsilon_{zz}-\epsilon_{xx})/\epsilon_{tot}$&$=n_e(\omega)-n_o(\omega)$&\\
\hline
\bf{\chb{Without vacancy}}&&&\\
\hline
Ordered LCMO &0.017&0.05& \\
Disordered LCMO &-0.149&-0.98& \\
Ordered LSCMO &-0.068&-0.63& \\
Disordered LSCMO&0.201&1.28&\\
\hline
\bf{\chb{With vacancy}}&&&\\
\hline
Ordered LCMO  &0.164&0.79& \\
Disordered LCMO  &-0.312&-1.97&  \\
Ordered LSCMO  &0.030 &0.21 & \\
Disordered LSCMO &0.541&2.11&\\
\hline
\hline
\end{tabular}
\label{t6}
\end{table} 
\section{Discussion}

In the case of ordered LCMO, we find that the oxidation states of Co and Mn ions are +2 and +4 respectively. However, with the introduction of anti-site disorder at the Co/Mn sites in LCMO, both Co and Mn ions are found to take up +3 oxidation state. The reason for this change is the following :
In the ordered state, Co$^{2+}$ (ionic radius = 0.89 \AA) alternates with Mn$^{4+}$ (ionic radius = 0.67 \AA), such that there are no Co-O-Co and Mn-O-Mn bonds. Such an arrangement of alternating larger (CoO$_6$) and smaller (MnO$_6$) octahedra helps in minimizing the lattice strain. However, in the disordered state, we have large number of Co-O-Co and Mn-O-Mn bonds, the result being an increase in the lattice strain. The large CoO$_6$ octahedra are forced to decrease their volume in order to accommodate the strain, which in-turn affects the d-orbital splitting. The energy difference between the t$_{2g}$ and e$_g$ orbitals increases as a result, which not only affects the Co spin-state but also its oxidation-state. The splitting (as calculated from the band centres of electronic density of states) is so large that (i) Co goes into low spin state (ii) Co-e$_g$ levels are raised above the Mn-e$_{g}$ levels (refer to the FIG. 8). As a result, after filling-in both the up-spin and down-spin levels of Co-t$_{2g}$ the last (seventh electron of Co$^{2+}$, coloured red in FIG. 8) goes to Mn-e$_{g}$ instead of Co-e$_g$, merely because of energetics. Therefore, Co and Mn both attain +3 oxidation state as a result of disorder at Co/Mn sites.

\begin{figure*}[htp]
\centering
\includegraphics[scale=.68]{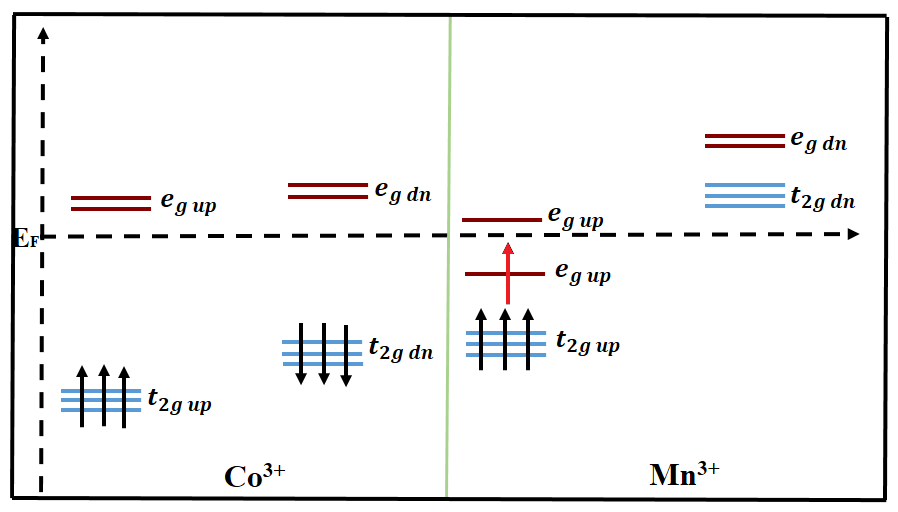}  
\caption{Energy level diagram of Co$^{3+}$ and Mn$^{3+}$ in disordered LaCoMnO$_6$. The red arrow depicts the electron which gets transferred from Co ion to Mn ion in the presence of anti-site disorder at Co/Mn sites.} 
\end{figure*}

\begin{figure*}[htp]
\centering
\includegraphics[scale=.80]{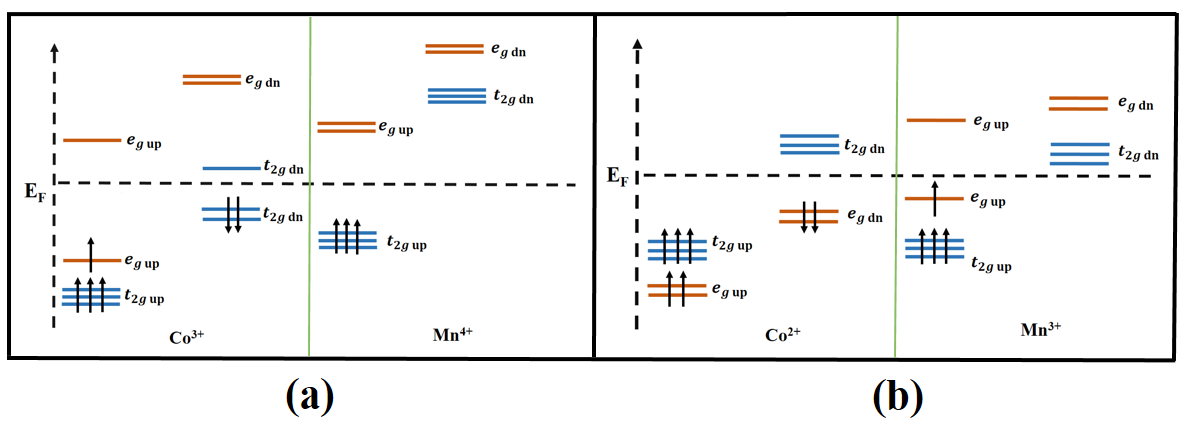}  
\caption{Energy level diagram of (a) Co$^{3+}$ and Mn$^{4+}$ in disordered LSCMO (b) Co$^{2+}$ and Mn$^{3+}$ in disordered LSCMO with O-site vacancy. }
\end{figure*}
 Electronic structure calculation of Sr doped LCMO shows that the charge imbalance caused by the introduction of Sr$^{2+}$ in place of La$^{3+}$ is taken care of by Co ions readjusting their charge and spin-states. Whereas, when oxygen vacancy is introduced into the system, it is the Mn ions that change their charge state to accommodate the charge imbalance. Therefore, in summary, Co accommodates the extra electron, whereas Mn accommodates the extra hole that is doped into the system. In an attempt to delineate the typical roles played by the two transition metal ions under varying circumstances, we perform an analysis of the relative energy levels of the Co/Mn-d orbitals in ordered/disordered LCMO/LSCMO. The energy level of the individual d-orbitals of Co and Mn ions were fixed at their corresponding band centres. Although this method is not expected to be quantitatively accurate, it serves to give a good qualitative understanding of the system nonetheless. As a representative case, we consider disordered LSCMO and disordered LSCMO containing O-site vacancy, the energy level diagram of the d-orbitals of the transition metal ions, whose degeneracy has now been lifted due to crystal field splitting, is shown in FIG. 9. For hole doping, a transition metal ion has to give away one electron. It is obvious that this electron will have to come from the highest filled level, which happens to be that of Co ion (left panel of FIG. 9). On the other hand, when an electron is doped into the system, this electron will go to the lowest available energy state. The right panel of FIG. 9 considers a situation of simultaneous electron and hole doping. Analysis of the energy level diagram shows Mn to have lowest unfilled level and Co to have the highest filled level. Therefore, Mn is expected to house the extra electron created by O-site vacancy. Whereas, Co should give away an electron as a response to hole doping. However, the PDOS and calculated moment of Co in disordered LSCMO is more Co$^{2+}$-like, instead of the expected Co$^{3+}$-like. The reason behind this, as already explained, is the strong co-valency prevailing in the system. The apparent anomalous behaviour of the crystal field splitting of Co-d orbitals in O-site-vacancy-containing LSCMO (FIG. 9(b)) can be attributed to the vacant O-site which is shared by two Co octahedra (refer to FIG. 1(h)). The absence of one anion in an octahedra is equivalent to a tetrahedral crystal field whose splitting is opposite to that of the octahedral case\cite{tetrahedral}. 

Antisite disorder at Co/Mn sites has been found to contribute to extrinsic magneto-dielectric effect at high temperature~\cite{R3c}. The intrinsic part of the magneto-dielectric effect, which arises at low temperatures due to asymmetric charge hopping between Co$^{2+\delta}$ to Mn$^{4-\delta}$, is found to be suppressed with rise in antisite disorder. Our calculations show that introduction of Sr increases anti-site disorder which is further enhanced with the induction of O-site vacancies. In the case of LCMO, electronic structure calculations show that anti-site disorder at Co/Mn sites introduces states at the fermi level in one of the spin-channels, thereby making it half-metallic. This half-metallicity, which is also evidenced in both ordered and disordered samples of Sr-doped LCMO (refer to TABLE II), is responsible for  the much celebrated collossal magnetoresistance~\cite{22} effect in these materials. However, when O-site vacancy is introduced, though there is an increase in anti-site disorder, the material is found to regain its insulating state. As already discussed, vacancy at an oxygen site leads to the disruption of the octahedral co-ordination of the affected CoO$_6$/MnO$_6$, thus reducing it to tetrahedral co-ordination. This removes parts of the \textit{d-p} hybridized bands from the fermi level, making the system insulating. 

In the presence of Sr doping, the values of the magnetic interaction parameters are found to be enhanced, whereas in the presence of O-site vacancy, their numerical values are reduced. In order to obtain an understanding of what causes the enhanced values of magnetic exchange interaction in the presence of Sr doping, we investigate the minor structural differences between ordered LCMO and LSCMO. Structural differences are mainly brought in by the size-difference between Sr$^{2+}$ and La$^{3+}$ and the corresponding effect that Sr has on the spin and charge state of Co which results in altered Co cationic radii. These modifications lead to the tailoring of the bond-lengths and bond-angles in the whole of the crystal lattice. As a result, the Co-O-Mn bond angle which was 156$^{\circ}$ in LCMO increases making the bond-angles closer to 180$^{\circ}$. It is to be noted that an ideal 180$^{\circ}$ bond angle would maximise the overlapping between orbitals giving rise to the most effective super-exchange mechanism. The bond-angles, Co-O-Mn in ordered LSCMO and Co-O-Mn/Co-O-Co/Mn-O-Mn in disordered LSCMO, end up being straightened to $\simeq$ 164$^{\circ}$ and $\simeq$ 175$^{\circ}$ (refer to TABLE S1 of SM\cite{supplemental}). The resultant bond-angles lead to more effective superexchange in Sr-doped system, which is reflected in the enhanced values of the 'J'-s. On the other-hand, oxygen vacancies cause a significant distortion in the structure, as deciphered from the calculated bond-angles and bond-lengths presented in TABLE S1 of the SM. The lower values of Co-O-Mn/Co-O-Co/Mn-O-Mn bond angles diminish the orbital overlap, thereby suppressing the super-exchange mechanism. Hence, the magnitude of magnetic exchange couplings are lower. The Curie temperature calculated with mean-field approximations is found to get reduced with the introduction of defects. The O-site vacancies have a much greater adverse impact on the $T_C$ than Sr-doping. This reduction is in response to the defect induced changes in the structure of LCMO as well as in the spin and charge state of the transition metal ions.    

However, defects are found to play a positive role as far as optical properties are concerned. We find that with electron and hole doping, LCMO (a material which is more or less optically isotropic) becomes highly anisotropic with significantly large values of birefringence, thereby making the defect-containing system a promising material for applications in optical devices. 
The two formula unit cell size that has been used for all our calculations may seem inadequate to capture the effect of antisite disorder at Co/Mn sites. We have therefore increased the cell size to 2$\times$1$\times$1 and carefully hand-picked few selected configurations of disordered LCMO and LSCMO where the (LaCoO$_3$)$_1$/(LaMnO$_3$)$_1$ kind of superlattice, which was seen in a single unit cell (refer to FIG. S1 of the SI\cite{supplemental}), does not arise. Our calculations suggest that the anti-site effect indeed persists in larger supercells. Refer to S11 of SM\cite{supplemental} for details. We have also validated our results for lower doping concentrations of Sr (refer to S12 of SM\cite{supplemental}) in LCMO.

\section{Conclusion} 
In this study we investigate the role played by defects in modulating the physical properties of LCMO. LCMO is found to crystallize in a monoclinic space group (\textit{P2$_1$/n}). However, with the introduction of Sr, LCMO appears in a mixed phase, having a dominant monoclinic \textit{P2$_1$/n} phase and a minor rhombohedral $R\overline{3}c$ phase. Sr$^{2+}$ doping at La$^{3+}$ site not only introduces charge imbalance, it also generates strain in the crystal. As a result, the Co$^{2+}$ ions take up a mixed valence state and Co/Mn octahedra distort significantly to accommodate the Sr ion. With the induction of O-site vacancy, however, it is the Mn ions that change their charge state. Moreover, the bond-lengths and bond-angles of the unit cell also alter to a great extent as a response to the disruption caused to the periodicity upon creation of oxygen vacancy. These modifications are expected to play a  pivotal role in deciding the physical properties of the material. For example, anti-site disorder at Co/Mn sites is found to get enhanced upon the introduction of defects. We find from our electronic structure calculation that Co/Mn anti-site disorder containing LCMO and ordered/disordered LSCMO is half-metallic. However, with vacancies at O-site, LCMO and LSCMO convert to an insulator. The Curie temperature estimated using a mean-field method is seen to reduce considerably with defects in the system. Whereas, the calculated linear optical properties suggest that defects introduce a significant anisotropy into the system. Therefore, we find that defects have their own pros and cons and depending upon the kind of application one is looking for, the defect concentration could be tuned to one's advantage. 
\section{Acknowledgment}
The authors acknowledge UGC and MoE for their financial support and appreciate the access to computing facilities of DST-FIST (phase-II) installed in department of physics at IIT Kharagpur.

\end{document}